\documentclass[aps, prb, amsmath, amssymb, reprint, superscriptaddress]{revtex4-1}

\newcommand*{\citen}[1]{%
  \begingroup
    \romannumeral-`\x
    \setcitestyle{numbers}%
    \cite{#1}%
  \endgroup   
}

\newcommand{\RNum}[1]{\uppercase\expandafter{\romannumeral #1\relax}}

\usepackage[dvipsnames]{xcolor}
\usepackage{graphicx}
\usepackage{dcolumn}
\usepackage{bm}

\begin{document}

\title[]{Instability in domain wall dynamics in almost compensated ferrimagnets}
\author{R. V. Ovcharov}
\affiliation{ 
Department of Physics, University of Gothenburg, Gothenburg 41296, Sweden
}

\author{B. A. Ivanov}
\affiliation{
Institute of Magnetism of NASU and MESU, Kyiv 03142, Ukraine
}
\affiliation{
William H. Miller III Department of Physics and Astronomy, Johns Hopkins University, Baltimore, MD 21218, USA.
}

\author{E. G. Galkina}
\affiliation{ 
Institute of Physics, National Academy of Sciences of Ukraine, Kyiv, 03028, Ukraine
}

\author{J. \AA kerman}
\affiliation{ 
Department of Physics, University of Gothenburg, Gothenburg 41296, Sweden
}
\affiliation{
Center for Science and Innovation in Spintronics, Tohoku University, Sendai 980-8577, Japan
}
\affiliation{
Research Institute of Electrical Communication, Tohoku University, Sendai 980-8577, Japan
}

\author{R. S. Khymyn}%
\affiliation{ 
Department of Physics, University of Gothenburg, Gothenburg 41296, Sweden
}

\date{\today}

\begin{abstract}
Nanoscale self-localized topological spin textures, such as domain walls and skyrmions, are of interest for the fundamental physics of magnets and spintronics applications. Ferrimagnets (FiMs), in the region close to the angular momentum compensation point, are promising materials because of their ultrafast spin dynamics at nonzero magnetization. In this work, we study specific features of the FiM domain wall (DW) dynamics, which are absent in both ferromagnets (FM) and antiferromagnets (AFM). In low-damping FMs and AFMs ($\alpha \ll 1$), the non-stationary forced motion of DWs is characterized by slow ($\mathit{t_{diss}}\propto 1/\alpha$) changes of the DW’s velocity and internal structure for all accepted values of the DW energy $E$ and its linear momentum $P$---a consequence of the stability of DWs for any value of $P$. In contrast, the dispersion law of FiM DWs has specific points, $P=P_{cr}$ and $E_{cr}=E(P_{cr})$, such that stable DWs are only present for $P<P_{cr}$, \emph{i.e.}, $P_{cr}$ and $E_{cr}$ act as endpoints in the $E(P)$ dependence. We show that when a field-like torque driven DW reaches this endpoint, it falls into a highly-non-equilibrium state with the excitation of fast ($t \ll \mathit{t_{diss}}$) and highly-nonlinear intra-wall magnetization dynamics, covering a wide frequency range up until the frequencies of propagating spin waves. The domain wall finally throws off an ``excessive'' energy by a short ``burst'' of the propagating spin waves and returns to the stationary state; the full picture of the forced motion is a periodic repetition of such ``explosive'' events.  
\end{abstract}

\maketitle

Domain walls (DWs) in magnetically ordered materials are transition regions of nanometer width that connect wide domains with different orientations of spins. They possess a number of interesting features, such as topological stability and the possibility of controlled manipulation including implementing high-speed motion. DWs are of great interest for the fundamental physics of magnets, as well as for various applications, such as high-density information storage\cite{chappert2007emergence, parkin2008magnetic} and logic\cite{allwood2005magnetic, luo2020currentdriven} devices. For these applications, the forced motion of a DW under the action of a relatively weak driving force (applied magnetic field~\cite{jing2022fielddriven} or spin torque~\cite{jing2024currentdriven}) is of the main interest.  From the viewpoint of fundamental physics of magnetism, these spin textures can be treated as magnetic solitons, important objects in the context of highly nonlinear spin dynamics. Solitonic characteristics of the DW moving with the velocity $v$, in particular, its structure and the maximum value of its velocity $v_c$, are usually treated for dissipation-free limit\cite{kosevich1990magnetic}. Here, the important characteristics are the dependencies of the DW integrals of motion, its energy $E$, and its linear momentum $P$, on the DW velocity.

The results obtained in the dissipation-free limit, allow the construction of a simple and universal description of the forced motion of the DW for the case of interest, weak dissipation, when the dissipative constant $\alpha \ll 1$ is small (usually $\alpha <1 0^{-2}$), and, accordingly,  non-small velocity of the DW till $v_c$ can be reached at the small value of the driving force, proportional to $\alpha$.
This approximate description is based on the perturbation theory for solitons using the collective variables approach~\cite{kivshar1989dynamics, tretiakov2008dynamics, kim2014propulsion, ovcharov2022spin, ovcharov2023antiferromagnetic, kim2023mechanics}. 

Within this approach, for non-stationary motion of the DW, the rate of change of its parameters (energy, velocity, etc) is again of the order of $\alpha$. If these conditions are met, the change in momentum is described by the following equation
\begin{equation}
\label{eq:motion}
    \frac{dP}{dt} = F_{fr} + F,
\end{equation}
where $F_{fr}$ and $F$ are the viscous friction and driving forces, respectively. Usually, the friction force is proportional to $\alpha vE(v)$. For non-stationary motion, the DW's momentum changes according to the Eq.~\eqref{eq:motion}, and all DW's parameters, the energy and the velocity $v=dE(P)/dP$, follow the dependence $E(P)$ (the DW's dispersion law) found in the dissipation-free limit. 

The velocity of the stationary motion $d_t P = 0$ can be found from the balance of forces, $F_{\mathrm{fr}} + F=0$. However, for many magnets, the energy of the DW moving with any possible DW velocity $v \leq v_c$, and, consequently, the frictional force, is limited from above. If the driving force exceeds this maximal value of frictional force, the DW momentum increases without limit, and the driven DW motion manifests nontrivial features determined exclusively by its dispersion relation, which is extremely different for different magnets.

In FMs, the spin dynamics is described by the famous Landau-Lifshitz equation\cite{landau1992theory} for magnetization $\mathbf{M}$ or, equivalently, the spin density $\mathbf{S}$, which are related as $\mathbf{M} = - g \mu_{B}\mathbf{S}$, $|\mathbf{M}|=M_0$, where $g$ is the Land\'e factor ($g$-factor), and $\mu_B$ is the Bohr magneton. This equation is determined by the Lagrangian that contains a nonanalytic term $\mathbf{A}\partial \mathbf{S}/\partial t$ with the singular vector function, vector-potential of the Dirac monopole field  $\mathbf{A}(\mathbf{S})$ (see, \emph{e.g.}, the monograph \cite{fradkin2013field}). As a consequence, the dependence $E(P)$ is periodic~\cite{galkina2000dispersion, tchernyshyov2015conserved}, $E_{FM}(P + 2P_{0,FM}) = E_{FM}(P)$  with the universal period $2P_{0,FM}= 2M_0/ \gamma= 2 \pi \hbar S$ per one atomic chain, perpendicular to the DW plane, where $\hbar$ is Planck’s constant, and $\gamma=g\mu_B/\hbar$ is the gyromagnetic ratio.

\begin{figure}[hbt!]
\includegraphics[width=0.45\textwidth]{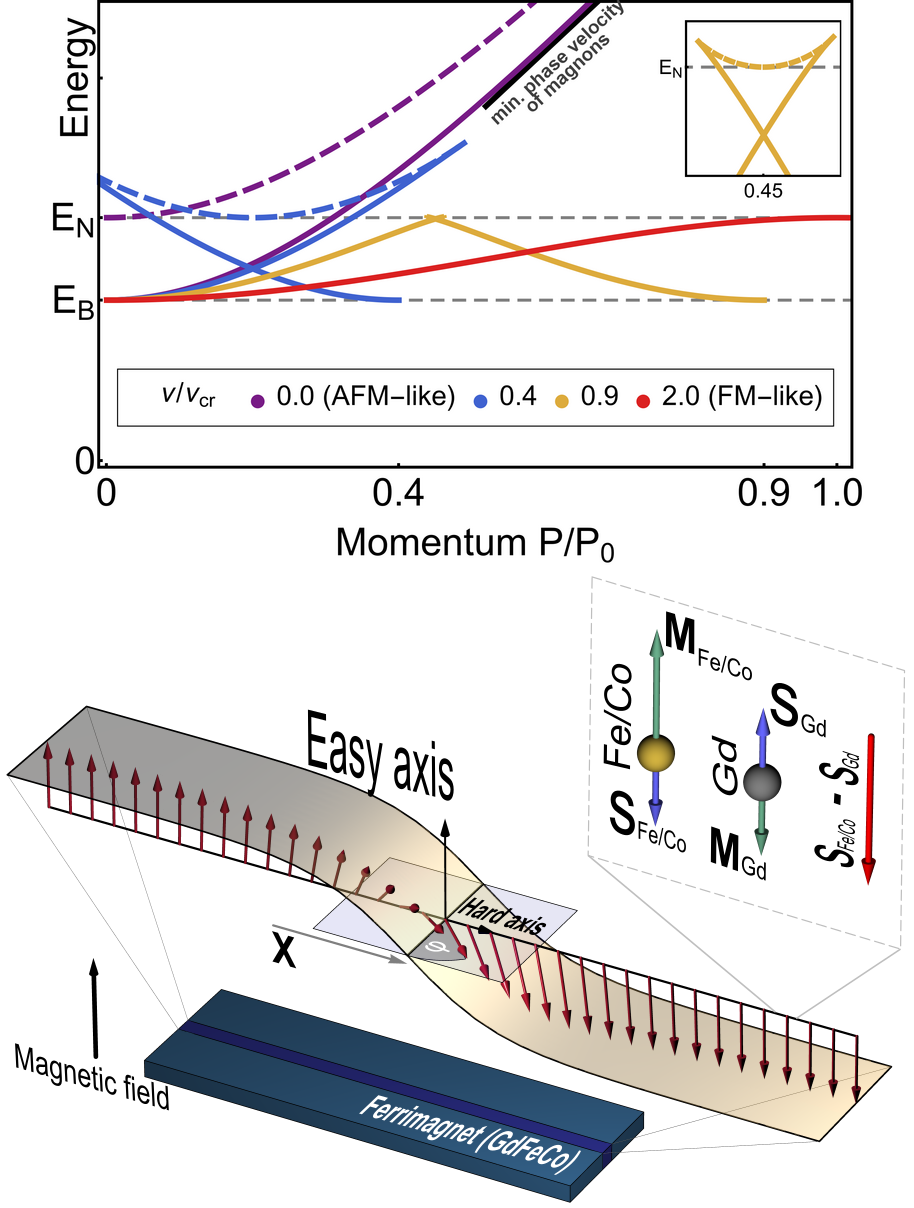}
 \caption{Different DW dispersion laws depending on the spin imbalance parameter of a FiM. Solid and dashed lines represent stable and unstable branches, respectively. Below, the illustration of a GdFeCo FiM sample is depicted, where a scaled strip shows the N\'eel vectors' orientation (red arrows) forming a DW. The external magnetic field is applied along the easy axis of the FiM, creating a driving force for the DW. $X$ and $\phi$ show DW collective coordinates, the position of its center, and the rotational plane angle inside, respectively. The inset shows the magnetic configuration of the GdFeCo cell at the AMCP.
 }
\label{img:schema}
\end{figure}
 
This type of dispersion law is also present for FiMs, see Fig.~\ref{img:schema}, where other possible types of dispersion laws for FiMs, including ``Lorentz''-invariant or containing aforementioned ending point, are drawn. 

The limit velocity of free motion of an FM DW, $v_{c, FM}$, is proportional to some relativistic constants, the magnetic in-plane anisotropy, or the magnetic dipole interaction; $v_{c, FM}$ vanishes in the exchange approximation. The DW energy and the value of the friction force are limited from above, $F_\mathrm{fr} \leq F_c$. If the driving force for the FM DW exceeds this maximal value, a nontrivial effect appears: a non-monotonic motion in response to constant driving field, the so-called ``Walker breakdown'', which is well known for decades\cite{schryer1974motion}. This effect is a direct consequence of the $E(P)$ periodicity, which is formally common to that for Bloch electrons, and, despite the different physical origins of the periodicity, the effect on DWs can be treated as an analogy of Bloch oscillations for electrons moving in a crystal under an action of a strong electric field. 

For AFMs, the character of spin dynamics is entirely different. For a two-sublattice AFM the spin dynamics can be described by a closed equation, the so-called $\sigma$\nobreakdash-model equation, for the single dynamical variable, the normalized N\`eel vector $\mathbf{l} = (\mathbf{S}_1 - \mathbf{S}_2)/2S$, where $S$ is the spin density of one of the sublattices. Here, the singularities present in the equations for $\mathbf{S}_{1,2}$ cancel each other, and the $\sigma$-model equation possesses formal ``Lorentz'' invariance with the chosen speed $c$, which equals the minimal phase velocity of magnons and is determined by exchange interaction only, see \emph{e.g.},  \cite{baryakhtar1985dynamics,kosevich1990magnetic}. This value has a sense of a limiting velocity of moving DWs. The dependencies of the freely moving DW on its velocity or momentum are the same as for the usual particle in relativistic mechanics, $E_\mathrm{AFM}(v)=E_{0}/\sqrt{1-v^2/c^2}$  or $E_\mathrm{AFM}(P)=\sqrt{E^2_{0}+c^2 P^2}$, where $E_{0}$ is the ``rest energy'' of a DW. Thus, for AFM DW, the energy and, consequently, the friction force grow infinitely as $v \to c$. For this reason, the velocity of the steady-state motion of the AFM DW driven by a constant force monotonously increases with the increasing force, approaching the limit value $c$, that was demonstrated experimentally for many AFMs, \emph{e.g.}, orthoferrites \cite{baryakhtar1985dynamics}. The value of $c$ is quite large, typically tens of km/s (20~km/s for orthoferrites), making AFM DWs suitable for high-speed spintronic applications\cite{gomonay2016high, yang2015domainwall, shiino2016antiferromagnetic, ovcharov2024emission}.   

A more common class of magnetic materials includes ferrimagnets (FiMs), which have magnetic sublattices with AFM exchange coupling and antiparallel orientation of the sublattice spins in the ground state. Contrary to AFMs, the sublattices are not equivalent, in particular, they can have non-equal spin densities $S_1 \neq S_2$, ($S_{1,2}=|\mathbf{S}_{1,2}|$), different g-factors, etc., see \emph{e.g.}, Ref.~\citen{ivanov2019ultrafast}. For a significant difference of sublattice spins, the spin dynamics in FiMs is the same as for standard FMs; otherwise, for the case of the exact balance of sublattice spins, $S_{1}=S_{2}$ FiMs resemble AFMs. As has been mentioned above, the non-trivial feature---the presence of an endpoint in the dispersion law of the DW---is unique for FiMs and has no analogues for FMs or AFMs \cite{galkina2019limiting}. 

 Let us compare the properties of FiMs and AFMs in more detail. The principal difference between FiMs and AFMs is that magnetic sublattices of a FiM consist of nonequivalent magnetic ions (or identical ions but in different crystallographic positions); as a consequence, they generally have non-equal spin densities $S_1 \neq S_2$ ($S_i=|\mathbf{S}_i|$) and can have distinct $g$-factors. The temperature dependencies of $S_1, \ S_2$ can also be different, allowing two compensation points in the ground state: the magnetization compensation point, where the net magnetization is zero, $\mathbf{M}_1+\mathbf{M}_2=0$, and the angular momentum compensation point (AMCP), characterized by a zero net spin density $\mathbf{S}_1+\mathbf{S}_2=0$. At the AMCP, where the spin densities of sublattices coincide, the spin dynamics of FiMs is, in fact, antiferromagnetic, and all AFM features, including the aforementioned exchange enhancement of the DW limiting velocity and Lorentz-invariance of the DW dynamics, are manifested\cite{ivanov2019ultrafast}. Thus, by adjusting the spin compensation of two sublattices either by the temperature\cite{stanciu2006ultrafast, ostler2011crystallographically} or the material composition\cite{kato2008compositional, bainsla2022ultrathin}, the limiting velocity of DWs can be substantially increased\cite{kim2017fast, blasing2018exchange,cai2020ultrafast, siddiqui2018currentinduceda, caretta2018fast, ghosh2021currentdriven, okuno2020spintransfer, zvezdin2020anomalies, logunov2021domain}. FiM thin films are also more easily fabricated than crystalline AFMs; even amorphous alloys like the famous Gd$_x$(FeCo)$_{1-x}$ can have FiM ordering with the AMCP about $x=0.25$. In addition, due to the aforementioned difference in $g$-factors, FiMs can have a nonzero magnetization at the AMCP, simplifying the control of DWs by a magnetic field. It is, therefore, not surprising that much attention has been paid to these materials in recent years. Far from the AMCP, the spin dynamics is practically the same as those of FMs. However, it turns out that in the most interesting region of close vicinity to the AMCP, where $S_1\neq S_2$, but $\nu=(S_1-S_2)/(S_1+S_2) \ll 1$, the dynamic properties of the DWs are much more intriguing than for both limit cases, FMs or AFMs. The analysis of the DW dispersion law for a simple model demonstrates the presence of quite a nontrivial element, the so-called end point at some finite value of the momentum\cite{galkina2019limiting}. Such behavior should manifest itself in the unique characteristics of the forced motion of the DWs in nearly-compensated FiMs, which is absent for both FMs and AFMs.

The nonlinear spin dynamics of a two-sublattices FiM not far from AMCP can be described in the way, common to AFMs, in terms of the AFM N\`eel vector $\mathbf{l}$ and normalized net angular momentum vector $\mathbf{s}=(\mathbf{S}_1+\mathbf{S}_2)/S_{tot}$. The vectors $\mathbf{l}$ and $\mathbf{s}$ satisfy the relations $\mathbf{s}\cdot \mathbf{l}=\nu$ and $\mathbf{s}^2 +  \mathbf{l}^2=1+\nu^2$, where $\nu=(S_{1}-S_{2})/(S_{1}+S_{2})$ defines the value of spin imbalance. In close proximity to the AMCP $\nu \ll 1$, the quantities $\nu$ and $s=|\mathbf{s}|$ are small, and the vector $\mathbf{l}$ can be treated as unit. Within this approach, $\mathbf{s}$ plays the role of a slave variable; it can be expressed through $\mathbf{l}$ and $\partial \mathbf{l}/\partial t$. Finally, the ferrimagnetic spin dynamics can be described by an equation that contains only the N\`eel vector $\mathbf{l}$, the so-called generalized $\sigma$-model equation\cite{ivanov2019ultrafast}. Since  $|\mathbf{l}|=1$, it is convenient to write down the  N\`eel vector in angular variables $\mathbf{l} = \{ \sin \theta \cos \phi, \sin \theta \sin \phi, \cos \theta\}$.

This generalized sigma-model equation can be written by variation of the Lagrangian $L=\int d \mathbf{r} \mathcal{L}[\mathbf{l}]$ and dissipation function $Q$. For the dissipation function, we will choose the simplest Gilbert form $Q=\alpha \hbar S_{tot} /2\int (\partial \mathbf{l}/\partial t)^2 d\mathbf{r}$. 

The Lagrangian density may be decomposed as $\mathcal{L}[\mathbf{l}]=T+G-W$. The kinetic part has the same form as for ``pure'' AFMs, $T=(A/2c^2)(\partial_t \mathbf{l})^2$, where $c=\sqrt{A\omega_{ex}/\hbar S_{tot}}$ is the characteristic speed, which coincides with the minimal phase velocity of spin waves at the AMCP $\nu = 0$, $\omega_{ex}= \gamma H_{ex}$ is the frequency defined by the exchange field $H_{ex}$, and $A$ is the constant of nonuniform exchange \cite{ivanov2019ultrafast}. 

The gyroscopic term $G[\mathbf{l}]$, linear over first time derivative of the vector $\mathbf{l}$, has the same structure as the dynamical term for the Landau-Lifshitz equation, $G=- \hbar (S_1-S_2) \mathbf{A} \partial_t \mathbf{l}$, where $\mathbf{A} = \mathbf{A}(\mathbf{l})$ is singular vector function, vector-potential of the Dirac monopole field with a unit magnetic charge, $\mathrm{rot}_{\mathbf{l}} \mathbf{A}=\mathbf{l}$; it differs from that for FM by replacing $S \to S_1-S_2$ \cite{ivanov2019ultrafast}. 

Finally, $W[\mathbf{l}] = (A/2) (\nabla {\rm {\bf l}})^2 + w_a (\mathbf{l})$ is the static energy density of the FiM, written through the vector $\mathbf{l}$ only and including the nonuniform exchange energy with the constant $A$ and anisotropy energy $w_a$.

The knowledge of the Lagrangian allows us to write down the expressions for the energy and linear momentum of the DW
\begin{subequations}
\label{eq:EP}
\begin{gather}
    E =\int d \mathbf{r} (T+W), \\
    \mathbf{P} = \int d \mathbf{r} [\hbar (S_1-S_2) \mathbf{A} \cdot \nabla{\mathbf{l}}-(A/c^2)(\partial_t \mathbf{l})\nabla{\mathbf{l}}].
\end{gather}
\end{subequations}

A qualitative analysis of the DW dispersion law can be done using a general traveling wave ansatz $\mathbf{l}=\mathbf{l}(\xi)$,  $\xi = x - vt$. The analysis for a FiM with non-zero spin imbalance, in this case, can be carried out on the basis of the known FM solution by substituting $\xi \rightarrow \xi / \sqrt{1-v^2/c^2}$ and reassigning the coefficients of the gyroscopic term, which is valid for an arbitrary form of anisotropy energy, see Ref.~\citen{galkina2022solitons} for more details. In the following, however, we consider the peculiarities arising for the FiM, limiting ourselves to the simplest form of the biaxial anisotropy energy, which in the angular variables reads
\begin{equation}\label{eq:wa}
    w_a = \frac{1}{2}K \sin^2 \theta \left( 1 + \rho \sin^2 \phi \right),
\end{equation}
where $K>0$ is the constant of the easy-axis anisotropy (expressed below through characteristic frequency $\hbar \omega_a = K/S_{tot}$) and the parameter $\rho$ determines the anisotropy in the basal plane $K_b = \rho K$. Here, the axis $z$ is chosen to be the easy axis of the FiM so that the ground state corresponds to $\theta = 0, \pi$.

For this model, an exact analytical solution for a FM DW, in the form of $\phi = \text{const}$ and $\theta = \theta(\xi)$, was constructed by Walker~\cite{dillon1963domains}. The corresponding generalization to the FiM case reads~\cite{ivanov1983nonlinear}
\begin{subequations}

\label{eq:dwProfile}
\begin{gather}
    \label{eq:dwProfileTheta}
     \cos \theta =  \tanh \left[ \frac{ \xi}{ l_0(v) }\right], \quad l_0(v) = \sqrt{\frac{A}{K}} \sqrt{\frac{1-v^2/c^2}{1+\rho \sin^2 \phi}}, \\
   \frac{v}{\sqrt{1- v^2/c^2}} = \frac{\rho}{\nu} \frac{\sqrt{AK}}{\hbar S_{tot}} \frac{ \sin \phi \cos \phi}{\sqrt{1+\rho \sin^2\phi}}. \label{eq:dwProfilePhi}
\end{gather}
\end{subequations}
Here, $l_0$ has the meaning of the DW width, and the $v$-dependence of $\phi$ is determined by Eq.~\eqref{eq:dwProfilePhi}. For the DW in the rest $v=0$, the values of $\phi$ are $0, \ \pi/2, \ \pi $, etc., and the characteristic scale $l_0$ is determined by the interplay between constants of nonuniform exchange $A$ and anisotropy $K$. If the vector $\mathbf{l}$ inside the wall turns in the energetically favorable plane $zx$, i.e., $\phi_B=0,\pi$, such a DW is referred to as a Bloch wall. Otherwise, for non-favorable plane $zy$ (the angle $\phi_N=\pi/2$ or $3\pi/2$), the DW is called a N\'eel wall. The singularity in Eq.~(\ref{eq:dwProfilePhi}) at $\nu=0$ indicates that velocity $v$ and angle $\phi$ are decoupled exactly at the AMCP, where $\phi=\phi_B$ or $\phi_N$ for any velocity $v<c$.

The expressions for energy and momentum of the moving FiM DW with the use of Eqs.~\eqref{eq:EP} were obtained in Ref.~\citen{galkina2019limiting}:
\begin{subequations}
\label{eq:dispersionWithVandPhi}
\begin{eqnarray}
    E(v, \phi) &=& E_0 \frac{\sqrt{1 + \rho \sin^2 \phi}}{\sqrt{1 - v^2 /c^2}}, \label{eq:energyWithVandPhi} \\
    P(v, \phi) &=& 2 \nu \hbar S_{tot} \phi +\frac{v}{c^2}E(v, \phi) \label{eq:momentumWithVandPhi}.
\end{eqnarray}
\end{subequations}

Dependencies $E(P)$ constructed using Eq.~(\ref{eq:dwProfilePhi}) are shown above in Fig.~\ref{img:schema} for different values of spin imbalance. DW momentum $P$ is the sum of two terms, which are characteristic of the FM and relativistic AFM contributions, respectively. Considering only the first term, the energy is a periodic function of the momentum, which is typical for FMs. This property is based on Walker's result, which states that spatial rotation of the parameter order within the DW occurs as a planar rotation (i.e., within a plane defined by $\phi=\text{const}$) at all velocities. In fact, even for more general models where this is not the case and $\phi=\phi(\xi)$, the periodicity holds for angles corresponding to the minimum and maximum of anisotropy energy with respect to $\phi$. In the biaxial case, these angles are $\phi = 0, \pi/2, \pi$, and so on. Since $0$ and $\pi$ are equivalent, this result remains valid for arbitrary biaxial anisotropy energy; for more details, see Ref.~\citen{galkina2022solitons}.

Considering both terms in the DW momentum, which is a feature of FiMs, makes the analysis more complicated. It turns out that while the formal periodicity of $E(P)$ is preserved, at a small but finite imbalance $\nu = \nu_{cr}=\sqrt{\rho \omega_{a}/ \omega_{ex}}$, the form of dispersion law changes qualitatively. The change in momentum is non-monotonic when $\nu<\nu_{cr}$, and a separate branch in the dispersion law is formed, which has been shown to be unstable~\cite{galkina2019limiting}; therefore, the spectrum of stable DW motion has an endpoint. This endpoint can be reached when the DW is forced to move in a non-stationary regime. Indeed, the friction force derived from the given dissipation function can be written using the energy of the moving DW, as well as its velocity~\cite{ivanov2020nonstationary}:
\begin{equation}
\label{eq:frictionForce}
F_{fr}(v) = -\alpha \frac{\hbar S_{tot}}{A} v E(v).
\end{equation}

For any nonzero spin imbalance, the wall energy is limited from above since Eq.~(\ref{eq:dwProfilePhi}) implies a maximum velocity $v_c$ that is less than $c$. As a result, the friction force is also limited $F_{max}= \alpha \rho K / \nu$. When the driving force exceeds the maximum value of the friction force, the wall momentum should increase monotonically according to Eq.~(\ref{eq:motion}), and following the characteristic dispersion law, the wall should ``fly off'' from it. When dissipation is small $\alpha \ll 1$, the maximum friction force $F_{max}$ is also small. However, once the DW reaches the endpoint, a sudden change in dynamic parameters is expected, violating the approximation of the soliton perturbation theory approach\cite{kivshar1989dynamics}.

\begin{figure}[t]
\includegraphics[width=1\linewidth]{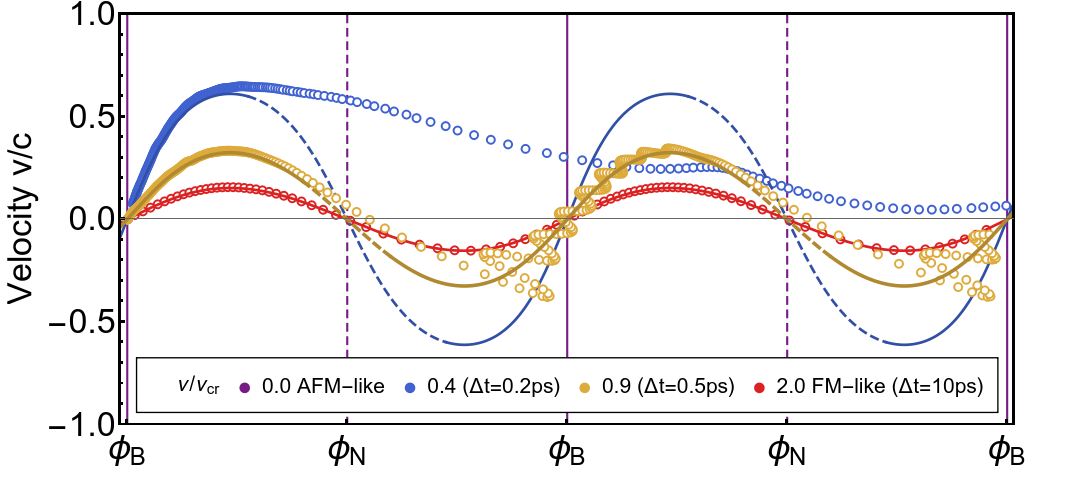}
 \caption{
 \textbf{Results of micromagnetic simulations.} DW velocity \textit{vs.} angle $\phi$ over time. Circles show the results of micromagnetic simulations, while lines are built according to analytical dependence (\ref{eq:dwProfilePhi}), where dashed segments indicate unstable regions found from the analysis of the dispersion laws. The applied force equals $2F_{max}$ for a given imbalance value, and therefore, the results are absent for the AFM-like case in which the nonstationary motion is not realized.
 }
\label{img:result}
\end{figure}

\begin{figure*}[hbt!]
\includegraphics[width=1\textwidth]{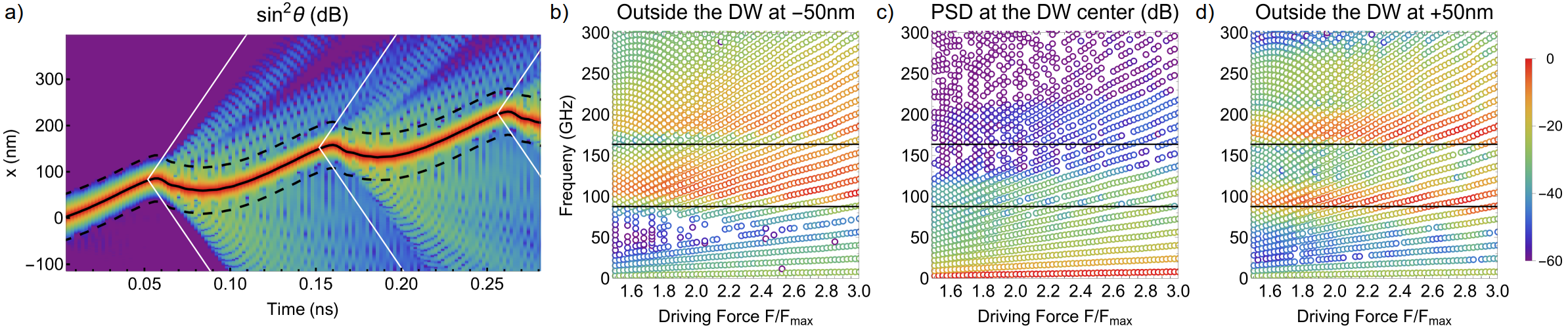}
 \caption{ \textbf{Spin waves emission after the DW ``explosion''.} (a) Shows the spatiotemporal evolution of the magnetization along the film. The white lines show the propagation with the maximum speed $c$ drawn from the points at which the angle $\phi$ inside the wall reaches the critical value. The solid black line follows the trajectory of the DW center, while the dashed black lines are $\pm50$~nm away. Dependencies on the driving force of the power spectral density calculated for the N\'eel vector following these trajectories are shown in (b), (c), and (d). The dots show the peaks of the spectra above the background. Horizontal black lines indicate resonance frequencies.  }
\label{img:sw}
\end{figure*}

To examine the DW dynamics when it reaches the regions of instability, we resort to micromagnetic simulations using the \emph{MuMax3} program\cite{vansteenkiste2014designa} (See Supplementary materials \cite{[{See Supplemental Material at }][{, which include Ref.~\citen{jing2022fielddriven, ivanov2020nonstationary}, for the details of micro-magnetic simulations and description of supplemental videos.}]supplemental}). The selected FiM corresponds to a Gd$_{25}$(FeCo)$_{75}$ alloy with $c=5.4$~km/s, $\omega_{ex}/2\pi =5$~THz, $K=100$~kJ/m$^3$ ($\omega_{a}/2\pi=2.8$~GHz), $\rho=1/2$, and $\alpha=10^{-3}$. The cases of the AMCP and imbalances of 0.4, 0.9, and 2.0 relative to the critical value $\nu_{cr}$ are inspected. With the help of micromagnetic simulations, it is also informative to visually inspect the changes, or lack thereof, that occur when a DW moves. For a better understanding, we provide supplementary videos showing the spatiotemporal evolution of the N\'eel vector during the DW motion. The cases of stationary motion and non-stationary regime for the FM-like scenario are shown in supplementary videos S1-3 and described in the supplementary materials. 

The cases of interest are considered for the spin imbalances of $0.9\nu_{cr}$ and $0.4\nu_{cr}$, chosen based on the characteristic placement of stable branch endpoints. At $\nu=0.9\nu_{cr}$, the momentum value at which the DW enters the instability region $P_{cr}$ remains smaller than the period $P_0$, $P_{cr}< P_0$. In Supplementary video S4, the result of simulations with an applied force of $2F_{max}$ is presented. Initially, the wall appears stable, resembling a ferromagnetic case, and the connection $v(\phi)$ given by Eq.~(\ref{eq:dwProfilePhi}) is maintained, see Fig.~\ref{img:result}. 
However, when the endpoint is reached, the angle inside the wall increases rapidly. Although the wall experiences negative velocities for some time, the motion is described by fast oscillations around analytical dependence $v(\phi)$ that eventually dampens. The wall manages to return to a stable configuration before the next critical point, after which the process repeats. 
The rapid passage of the instability region leads to high-frequency excitations that change the structure of the wall and can spread beyond its boundaries, see Fig.~\ref{img:sw}~(a). To analyze them, we computed the fast Fourier transform of the N\'eel vector at different places in a frame of reference moving along with the wall. Fig.~\ref{img:sw}~(b-d) shows that the spectrum at the DW center resembles a frequency comb in a wide frequency range. Those excitations below the resonance frequencies quickly decay in evanescent modes, but those with higher frequencies propagate as spin waves far away from the DW center.

As the imbalance further decreases, the energy and momentum of the DW inside the instability region are much higher, \emph{e.g.}, the value of $P_{cr}$ can even overcome the period $P_{0}$; see respective dispersion law in Fig.~\ref{img:schema}.  The simulations with $\nu=0.4\nu_{cr}$ show that the wall initially behaves similarly to the previous case, but after the ``jump'', the wall is not captured by the first stable region. The features of the wall movement are stochastic; the dynamics of the angle $\phi$ may not be attracted to the stable part of $v(\phi)$ at all, see Fig.~\ref{img:result}. In Supplementary video S5, it is evident that the wall only experiences a slowdown during the ``explosions''; the moments when it moves back are barely noticeable.

In conclusion, we have demonstrated that the forced motion of the DW in FiM near the AMCP exhibits a unique behavior that is absent for FMs or AFMs. In particular, the non-stationary DW motion is accompanied by a ``splash of excess energy'' when it reaches the endpoints of its dispersion law. Such periodic ``explosions'' are described by the excitation of fast intra-wall dynamics leading to the emission of propagating spin waves. This emission relaxes the DW back to the stable branch if the period between falling into the highly non-equilibrium state is long enough. To our knowledge, this specific scenario has not been studied before.

This project is partly funded by the European Research Council (ERC) under the European Union’s Horizon 2020 research and innovation programme (Grant TOPSPIN No 835068) and the Swedish Research Council Framework Grant Dnr. 2016-05980.

\bibliography{main}

\begin{thebibliography}{44}%
\makeatletter
\providecommand \@ifxundefined [1]{%
 \@ifx{#1\undefined}
}%
\providecommand \@ifnum [1]{%
 \ifnum #1\expandafter \@firstoftwo
 \else \expandafter \@secondoftwo
 \fi
}%
\providecommand \@ifx [1]{%
 \ifx #1\expandafter \@firstoftwo
 \else \expandafter \@secondoftwo
 \fi
}%
\providecommand \natexlab [1]{#1}%
\providecommand \enquote  [1]{``#1''}%
\providecommand \bibnamefont  [1]{#1}%
\providecommand \bibfnamefont [1]{#1}%
\providecommand \citenamefont [1]{#1}%
\providecommand \href@noop [0]{\@secondoftwo}%
\providecommand \href [0]{\begingroup \@sanitize@url \@href}%
\providecommand \@href[1]{\@@startlink{#1}\@@href}%
\providecommand \@@href[1]{\endgroup#1\@@endlink}%
\providecommand \@sanitize@url [0]{\catcode `\\12\catcode `\$12\catcode `\&12\catcode `\#12\catcode `\^12\catcode `\_12\catcode `\%12\relax}%
\providecommand \@@startlink[1]{}%
\providecommand \@@endlink[0]{}%
\providecommand \url  [0]{\begingroup\@sanitize@url \@url }%
\providecommand \@url [1]{\endgroup\@href {#1}{\urlprefix }}%
\providecommand \urlprefix  [0]{URL }%
\providecommand \Eprint [0]{\href }%
\providecommand \doibase [0]{http://dx.doi.org/}%
\providecommand \selectlanguage [0]{\@gobble}%
\providecommand \bibinfo  [0]{\@secondoftwo}%
\providecommand \bibfield  [0]{\@secondoftwo}%
\providecommand \translation [1]{[#1]}%
\providecommand \BibitemOpen [0]{}%
\providecommand \bibitemStop [0]{}%
\providecommand \bibitemNoStop [0]{.\EOS\space}%
\providecommand \EOS [0]{\spacefactor3000\relax}%
\providecommand \BibitemShut  [1]{\csname bibitem#1\endcsname}%
\let\auto@bib@innerbib\@empty
%</preamble>
\bibitem [{\citenamefont {Chappert}\ \emph {et~al.}(2007)\citenamefont {Chappert}, \citenamefont {Fert},\ and\ \citenamefont {Van~Dau}}]{chappert2007emergence}%
  \BibitemOpen
  \bibfield  {author} {\bibinfo {author} {\bibfnamefont {C.}~\bibnamefont {Chappert}}, \bibinfo {author} {\bibfnamefont {A.}~\bibnamefont {Fert}}, \ and\ \bibinfo {author} {\bibfnamefont {F.~N.}\ \bibnamefont {Van~Dau}},\ }\href {\doibase 10.1038/nmat2024} {\bibfield  {journal} {\bibinfo  {journal} {Nature Mater}\ }\textbf {\bibinfo {volume} {6}},\ \bibinfo {pages} {813} (\bibinfo {year} {2007})}\BibitemShut {NoStop}%
\bibitem [{\citenamefont {Parkin}\ \emph {et~al.}(2008)\citenamefont {Parkin}, \citenamefont {Hayashi},\ and\ \citenamefont {Thomas}}]{parkin2008magnetic}%
  \BibitemOpen
  \bibfield  {author} {\bibinfo {author} {\bibfnamefont {S.~S.~P.}\ \bibnamefont {Parkin}}, \bibinfo {author} {\bibfnamefont {M.}~\bibnamefont {Hayashi}}, \ and\ \bibinfo {author} {\bibfnamefont {L.}~\bibnamefont {Thomas}},\ }\href {\doibase 10.1126/science.1145799} {\bibfield  {journal} {\bibinfo  {journal} {Science}\ }\textbf {\bibinfo {volume} {320}},\ \bibinfo {pages} {190} (\bibinfo {year} {2008})}\BibitemShut {NoStop}%
\bibitem [{\citenamefont {Allwood}\ \emph {et~al.}(2005)\citenamefont {Allwood}, \citenamefont {Xiong}, \citenamefont {Faulkner}, \citenamefont {Atkinson}, \citenamefont {Petit},\ and\ \citenamefont {Cowburn}}]{allwood2005magnetic}%
  \BibitemOpen
  \bibfield  {author} {\bibinfo {author} {\bibfnamefont {D.~A.}\ \bibnamefont {Allwood}}, \bibinfo {author} {\bibfnamefont {G.}~\bibnamefont {Xiong}}, \bibinfo {author} {\bibfnamefont {C.~C.}\ \bibnamefont {Faulkner}}, \bibinfo {author} {\bibfnamefont {D.}~\bibnamefont {Atkinson}}, \bibinfo {author} {\bibfnamefont {D.}~\bibnamefont {Petit}}, \ and\ \bibinfo {author} {\bibfnamefont {R.~P.}\ \bibnamefont {Cowburn}},\ }\href {\doibase 10.1126/science.1108813} {\bibfield  {journal} {\bibinfo  {journal} {Science}\ }\textbf {\bibinfo {volume} {309}},\ \bibinfo {pages} {1688} (\bibinfo {year} {2005})}\BibitemShut {NoStop}%
\bibitem [{\citenamefont {Luo}\ \emph {et~al.}(2020)\citenamefont {Luo}, \citenamefont {Hrabec}, \citenamefont {Dao}, \citenamefont {Sala}, \citenamefont {Finizio}, \citenamefont {Feng}, \citenamefont {Mayr}, \citenamefont {Raabe}, \citenamefont {Gambardella},\ and\ \citenamefont {Heyderman}}]{luo2020currentdriven}%
  \BibitemOpen
  \bibfield  {author} {\bibinfo {author} {\bibfnamefont {Z.}~\bibnamefont {Luo}}, \bibinfo {author} {\bibfnamefont {A.}~\bibnamefont {Hrabec}}, \bibinfo {author} {\bibfnamefont {T.~P.}\ \bibnamefont {Dao}}, \bibinfo {author} {\bibfnamefont {G.}~\bibnamefont {Sala}}, \bibinfo {author} {\bibfnamefont {S.}~\bibnamefont {Finizio}}, \bibinfo {author} {\bibfnamefont {J.}~\bibnamefont {Feng}}, \bibinfo {author} {\bibfnamefont {S.}~\bibnamefont {Mayr}}, \bibinfo {author} {\bibfnamefont {J.}~\bibnamefont {Raabe}}, \bibinfo {author} {\bibfnamefont {P.}~\bibnamefont {Gambardella}}, \ and\ \bibinfo {author} {\bibfnamefont {L.~J.}\ \bibnamefont {Heyderman}},\ }\href {\doibase 10.1038/s41586-020-2061-y} {\bibfield  {journal} {\bibinfo  {journal} {Nature}\ }\textbf {\bibinfo {volume} {579}},\ \bibinfo {pages} {214} (\bibinfo {year} {2020})}\BibitemShut {NoStop}%
\bibitem [{\citenamefont {Jing}\ \emph {et~al.}(2022)\citenamefont {Jing}, \citenamefont {Gong},\ and\ \citenamefont {Wang}}]{jing2022fielddriven}%
  \BibitemOpen
  \bibfield  {author} {\bibinfo {author} {\bibfnamefont {K.~Y.}\ \bibnamefont {Jing}}, \bibinfo {author} {\bibfnamefont {X.}~\bibnamefont {Gong}}, \ and\ \bibinfo {author} {\bibfnamefont {X.~R.}\ \bibnamefont {Wang}},\ }\href {\doibase 10.1103/PhysRevB.106.174429} {\bibfield  {journal} {\bibinfo  {journal} {Phys. Rev. B}\ }\textbf {\bibinfo {volume} {106}},\ \bibinfo {pages} {174429} (\bibinfo {year} {2022})}\BibitemShut {NoStop}%
\bibitem [{\citenamefont {Jing}\ \emph {et~al.}(2024)\citenamefont {Jing}, \citenamefont {Sun},\ and\ \citenamefont {Wang}}]{jing2024currentdriven}%
  \BibitemOpen
  \bibfield  {author} {\bibinfo {author} {\bibfnamefont {K.~Y.}\ \bibnamefont {Jing}}, \bibinfo {author} {\bibfnamefont {Z.-Z.}\ \bibnamefont {Sun}}, \ and\ \bibinfo {author} {\bibfnamefont {X.~R.}\ \bibnamefont {Wang}},\ }\href {\doibase 10.1103/PhysRevB.110.054414} {\bibfield  {journal} {\bibinfo  {journal} {Phys. Rev. B}\ }\textbf {\bibinfo {volume} {110}},\ \bibinfo {pages} {054414} (\bibinfo {year} {2024})}\BibitemShut {NoStop}%
\bibitem [{\citenamefont {Kosevich}\ \emph {et~al.}(1990)\citenamefont {Kosevich}, \citenamefont {Ivanov},\ and\ \citenamefont {Kovalev}}]{kosevich1990magnetic}%
  \BibitemOpen
  \bibfield  {author} {\bibinfo {author} {\bibfnamefont {A.~M.}\ \bibnamefont {Kosevich}}, \bibinfo {author} {\bibfnamefont {B.~A.}\ \bibnamefont {Ivanov}}, \ and\ \bibinfo {author} {\bibfnamefont {A.~S.}\ \bibnamefont {Kovalev}},\ }\href {\doibase 10.1016/0370-1573(90)90130-T} {\bibfield  {journal} {\bibinfo  {journal} {Physics Reports}\ }\textbf {\bibinfo {volume} {194}},\ \bibinfo {pages} {117} (\bibinfo {year} {1990})}\BibitemShut {NoStop}%
\bibitem [{\citenamefont {Kivshar}\ and\ \citenamefont {Malomed}(1989)}]{kivshar1989dynamics}%
  \BibitemOpen
  \bibfield  {author} {\bibinfo {author} {\bibfnamefont {Y.~S.}\ \bibnamefont {Kivshar}}\ and\ \bibinfo {author} {\bibfnamefont {B.~A.}\ \bibnamefont {Malomed}},\ }\href {\doibase 10.1103/RevModPhys.61.763} {\bibfield  {journal} {\bibinfo  {journal} {Rev. Mod. Phys.}\ }\textbf {\bibinfo {volume} {61}},\ \bibinfo {pages} {763} (\bibinfo {year} {1989})}\BibitemShut {NoStop}%
\bibitem [{\citenamefont {Tretiakov}\ \emph {et~al.}(2008)\citenamefont {Tretiakov}, \citenamefont {Clarke}, \citenamefont {Chern}, \citenamefont {Bazaliy},\ and\ \citenamefont {Tchernyshyov}}]{tretiakov2008dynamics}%
  \BibitemOpen
  \bibfield  {author} {\bibinfo {author} {\bibfnamefont {O.~A.}\ \bibnamefont {Tretiakov}}, \bibinfo {author} {\bibfnamefont {D.}~\bibnamefont {Clarke}}, \bibinfo {author} {\bibfnamefont {G.-W.}\ \bibnamefont {Chern}}, \bibinfo {author} {\bibfnamefont {Y.~B.}\ \bibnamefont {Bazaliy}}, \ and\ \bibinfo {author} {\bibfnamefont {O.}~\bibnamefont {Tchernyshyov}},\ }\href {\doibase 10.1103/PhysRevLett.100.127204} {\bibfield  {journal} {\bibinfo  {journal} {Phys. Rev. Lett.}\ }\textbf {\bibinfo {volume} {100}},\ \bibinfo {pages} {127204} (\bibinfo {year} {2008})}\BibitemShut {NoStop}%
\bibitem [{\citenamefont {Kim}\ \emph {et~al.}(2014)\citenamefont {Kim}, \citenamefont {Tserkovnyak},\ and\ \citenamefont {Tchernyshyov}}]{kim2014propulsion}%
  \BibitemOpen
  \bibfield  {author} {\bibinfo {author} {\bibfnamefont {S.~K.}\ \bibnamefont {Kim}}, \bibinfo {author} {\bibfnamefont {Y.}~\bibnamefont {Tserkovnyak}}, \ and\ \bibinfo {author} {\bibfnamefont {O.}~\bibnamefont {Tchernyshyov}},\ }\href@noop {} {\bibfield  {journal} {\bibinfo  {journal} {Physical Review B}\ }\textbf {\bibinfo {volume} {90}},\ \bibinfo {pages} {104406} (\bibinfo {year} {2014})}\BibitemShut {NoStop}%
\bibitem [{\citenamefont {Ovcharov}\ \emph {et~al.}(2022)\citenamefont {Ovcharov}, \citenamefont {Galkina}, \citenamefont {Ivanov},\ and\ \citenamefont {Khymyn}}]{ovcharov2022spin}%
  \BibitemOpen
  \bibfield  {author} {\bibinfo {author} {\bibfnamefont {R.~V.}\ \bibnamefont {Ovcharov}}, \bibinfo {author} {\bibfnamefont {E.~G.}\ \bibnamefont {Galkina}}, \bibinfo {author} {\bibfnamefont {B.~A.}\ \bibnamefont {Ivanov}}, \ and\ \bibinfo {author} {\bibfnamefont {R.~S.}\ \bibnamefont {Khymyn}},\ }\href {\doibase 10.1103/PhysRevApplied.18.024047} {\bibfield  {journal} {\bibinfo  {journal} {Physical Review Applied}\ }\textbf {\bibinfo {volume} {18}},\ \bibinfo {pages} {024047} (\bibinfo {year} {2022})}\BibitemShut {NoStop}%
\bibitem [{\citenamefont {Ovcharov}\ \emph {et~al.}(2023)\citenamefont {Ovcharov}, \citenamefont {Ivanov}, \citenamefont {\AA{}kerman},\ and\ \citenamefont {Khymyn}}]{ovcharov2023antiferromagnetic}%
  \BibitemOpen
  \bibfield  {author} {\bibinfo {author} {\bibfnamefont {R.~V.}\ \bibnamefont {Ovcharov}}, \bibinfo {author} {\bibfnamefont {B.~A.}\ \bibnamefont {Ivanov}}, \bibinfo {author} {\bibfnamefont {J.}~\bibnamefont {\AA{}kerman}}, \ and\ \bibinfo {author} {\bibfnamefont {R.~S.}\ \bibnamefont {Khymyn}},\ }\href {\doibase 10.1103/PhysRevApplied.20.034060} {\bibfield  {journal} {\bibinfo  {journal} {Physical Review Applied}\ }\textbf {\bibinfo {volume} {20}},\ \bibinfo {pages} {034060} (\bibinfo {year} {2023})}\BibitemShut {NoStop}%
\bibitem [{\citenamefont {Kim}\ and\ \citenamefont {Tchernyshyov}(2023)}]{kim2023mechanics}%
  \BibitemOpen
  \bibfield  {author} {\bibinfo {author} {\bibfnamefont {S.~K.}\ \bibnamefont {Kim}}\ and\ \bibinfo {author} {\bibfnamefont {O.}~\bibnamefont {Tchernyshyov}},\ }\href@noop {} {\bibfield  {journal} {\bibinfo  {journal} {Journal of Physics: Condensed Matter}\ }\textbf {\bibinfo {volume} {35}},\ \bibinfo {pages} {134002} (\bibinfo {year} {2023})}\BibitemShut {NoStop}%
\bibitem [{\citenamefont {Landau}\ and\ \citenamefont {Lifshitz}(1992)}]{landau1992theory}%
  \BibitemOpen
  \bibfield  {author} {\bibinfo {author} {\bibfnamefont {L.}~\bibnamefont {Landau}}\ and\ \bibinfo {author} {\bibfnamefont {E.}~\bibnamefont {Lifshitz}},\ }in\ \href {\doibase 10.1016/B978-0-08-036364-6.50008-9} {\emph {\bibinfo {booktitle} {Perspectives in {{Theoretical Physics}}}}},\ \bibinfo {editor} {edited by\ \bibinfo {editor} {\bibfnamefont {L.~P.}\ \bibnamefont {Pitaevski}}}\ (\bibinfo  {publisher} {{Pergamon}},\ \bibinfo {address} {{Amsterdam}},\ \bibinfo {year} {1992})\ pp.\ \bibinfo {pages} {51--65}\BibitemShut {NoStop}%
\bibitem [{\citenamefont {Fradkin}(2013)}]{fradkin2013field}%
  \BibitemOpen
  \bibfield  {author} {\bibinfo {author} {\bibfnamefont {E.}~\bibnamefont {Fradkin}},\ }\href {\doibase 10.1017/CBO9781139015509} {\emph {\bibinfo {title} {Field {{Theories}} of {{Condensed Matter Physics}}}}},\ \bibinfo {edition} {2nd}\ ed.\ (\bibinfo  {publisher} {Cambridge University Press},\ \bibinfo {address} {Cambridge},\ \bibinfo {year} {2013})\BibitemShut {NoStop}%
\bibitem [{\citenamefont {Galkina}\ and\ \citenamefont {Ivanov}(2000)}]{galkina2000dispersion}%
  \BibitemOpen
  \bibfield  {author} {\bibinfo {author} {\bibfnamefont {E.~G.}\ \bibnamefont {Galkina}}\ and\ \bibinfo {author} {\bibfnamefont {B.~A.}\ \bibnamefont {Ivanov}},\ }\href@noop {} {\bibfield  {journal} {\bibinfo  {journal} {JETP Letters}\ }\textbf {\bibinfo {volume} {71}},\ \bibinfo {pages} {259} (\bibinfo {year} {2000})}\BibitemShut {NoStop}%
\bibitem [{\citenamefont {Tchernyshyov}(2015)}]{tchernyshyov2015conserved}%
  \BibitemOpen
  \bibfield  {author} {\bibinfo {author} {\bibfnamefont {O.}~\bibnamefont {Tchernyshyov}},\ }\href@noop {} {\bibfield  {journal} {\bibinfo  {journal} {Annals of Physics}\ }\textbf {\bibinfo {volume} {363}},\ \bibinfo {pages} {98} (\bibinfo {year} {2015})}\BibitemShut {NoStop}%
\bibitem [{\citenamefont {Schryer}\ and\ \citenamefont {Walker}(1974)}]{schryer1974motion}%
  \BibitemOpen
  \bibfield  {author} {\bibinfo {author} {\bibfnamefont {N.~L.}\ \bibnamefont {Schryer}}\ and\ \bibinfo {author} {\bibfnamefont {L.~R.}\ \bibnamefont {Walker}},\ }\href {\doibase 10.1063/1.1663252} {\bibfield  {journal} {\bibinfo  {journal} {Journal of Applied Physics}\ }\textbf {\bibinfo {volume} {45}},\ \bibinfo {pages} {5406} (\bibinfo {year} {1974})}\BibitemShut {NoStop}%
\bibitem [{\citenamefont {Bar'yakhtar}\ \emph {et~al.}(1985)\citenamefont {Bar'yakhtar}, \citenamefont {Ivanov},\ and\ \citenamefont {Chetkin}}]{baryakhtar1985dynamics}%
  \BibitemOpen
  \bibfield  {author} {\bibinfo {author} {\bibfnamefont {V.~G.}\ \bibnamefont {Bar'yakhtar}}, \bibinfo {author} {\bibfnamefont {B.~A.}\ \bibnamefont {Ivanov}}, \ and\ \bibinfo {author} {\bibfnamefont {M.~V.}\ \bibnamefont {Chetkin}},\ }\href@noop {} {\bibfield  {journal} {\bibinfo  {journal} {Uspekhi Fizicheskikh Nauk}\ }\textbf {\bibinfo {volume} {46}},\ \bibinfo {pages} {417} (\bibinfo {year} {1985})}\BibitemShut {NoStop}%
\bibitem [{\citenamefont {Gomonay}\ \emph {et~al.}(2016)\citenamefont {Gomonay}, \citenamefont {Jungwirth},\ and\ \citenamefont {Sinova}}]{gomonay2016high}%
  \BibitemOpen
  \bibfield  {author} {\bibinfo {author} {\bibfnamefont {O.}~\bibnamefont {Gomonay}}, \bibinfo {author} {\bibfnamefont {T.}~\bibnamefont {Jungwirth}}, \ and\ \bibinfo {author} {\bibfnamefont {J.}~\bibnamefont {Sinova}},\ }\href {\doibase 10.1103/PhysRevLett.117.017202} {\bibfield  {journal} {\bibinfo  {journal} {Phys. Rev. Lett.}\ }\textbf {\bibinfo {volume} {117}},\ \bibinfo {pages} {017202} (\bibinfo {year} {2016})}\BibitemShut {NoStop}%
\bibitem [{\citenamefont {Yang}\ \emph {et~al.}(2015)\citenamefont {Yang}, \citenamefont {Ryu},\ and\ \citenamefont {Parkin}}]{yang2015domainwall}%
  \BibitemOpen
  \bibfield  {author} {\bibinfo {author} {\bibfnamefont {S.-H.}\ \bibnamefont {Yang}}, \bibinfo {author} {\bibfnamefont {K.-S.}\ \bibnamefont {Ryu}}, \ and\ \bibinfo {author} {\bibfnamefont {S.}~\bibnamefont {Parkin}},\ }\href {\doibase 10.1038/nnano.2014.324} {\bibfield  {journal} {\bibinfo  {journal} {Nature Nanotech}\ }\textbf {\bibinfo {volume} {10}},\ \bibinfo {pages} {221} (\bibinfo {year} {2015})}\BibitemShut {NoStop}%
\bibitem [{\citenamefont {Shiino}\ \emph {et~al.}(2016)\citenamefont {Shiino}, \citenamefont {Oh}, \citenamefont {Haney}, \citenamefont {Lee}, \citenamefont {Go}, \citenamefont {Park},\ and\ \citenamefont {Lee}}]{shiino2016antiferromagnetic}%
  \BibitemOpen
  \bibfield  {author} {\bibinfo {author} {\bibfnamefont {T.}~\bibnamefont {Shiino}}, \bibinfo {author} {\bibfnamefont {S.-H.}\ \bibnamefont {Oh}}, \bibinfo {author} {\bibfnamefont {P.~M.}\ \bibnamefont {Haney}}, \bibinfo {author} {\bibfnamefont {S.-W.}\ \bibnamefont {Lee}}, \bibinfo {author} {\bibfnamefont {G.}~\bibnamefont {Go}}, \bibinfo {author} {\bibfnamefont {B.-G.}\ \bibnamefont {Park}}, \ and\ \bibinfo {author} {\bibfnamefont {K.-J.}\ \bibnamefont {Lee}},\ }\href {\doibase 10.1103/PhysRevLett.117.087203} {\bibfield  {journal} {\bibinfo  {journal} {Phys. Rev. Lett.}\ }\textbf {\bibinfo {volume} {117}},\ \bibinfo {pages} {087203} (\bibinfo {year} {2016})}\BibitemShut {NoStop}%
\bibitem [{\citenamefont {Ovcharov}\ \emph {et~al.}(2024)\citenamefont {Ovcharov}, \citenamefont {Ivanov}, \citenamefont {{\AA}kerman},\ and\ \citenamefont {Khymyn}}]{ovcharov2024emission}%
  \BibitemOpen
  \bibfield  {author} {\bibinfo {author} {\bibfnamefont {R.~V.}\ \bibnamefont {Ovcharov}}, \bibinfo {author} {\bibfnamefont {B.~A.}\ \bibnamefont {Ivanov}}, \bibinfo {author} {\bibfnamefont {J.}~\bibnamefont {{\AA}kerman}}, \ and\ \bibinfo {author} {\bibfnamefont {R.~S.}\ \bibnamefont {Khymyn}},\ }\href {\doibase 10.1103/PhysRevB.109.L140406} {\bibfield  {journal} {\bibinfo  {journal} {Physical Review B}\ }\textbf {\bibinfo {volume} {109}},\ \bibinfo {pages} {L140406} (\bibinfo {year} {2024})}\BibitemShut {NoStop}%
\bibitem [{\citenamefont {Ivanov}(2019)}]{ivanov2019ultrafast}%
  \BibitemOpen
  \bibfield  {author} {\bibinfo {author} {\bibfnamefont {B.~A.}\ \bibnamefont {Ivanov}},\ }\href {\doibase 10.1063/1.5121265} {\bibfield  {journal} {\bibinfo  {journal} {Low Temperature Physics}\ }\textbf {\bibinfo {volume} {45}},\ \bibinfo {pages} {935} (\bibinfo {year} {2019})}\BibitemShut {NoStop}%
\bibitem [{\citenamefont {Galkina}\ \emph {et~al.}(2019)\citenamefont {Galkina}, \citenamefont {Zaspel}, \citenamefont {Ivanov}, \citenamefont {Kulagin},\ and\ \citenamefont {Lerman}}]{galkina2019limiting}%
  \BibitemOpen
  \bibfield  {author} {\bibinfo {author} {\bibfnamefont {E.~G.}\ \bibnamefont {Galkina}}, \bibinfo {author} {\bibfnamefont {C.~E.}\ \bibnamefont {Zaspel}}, \bibinfo {author} {\bibfnamefont {B.~A.}\ \bibnamefont {Ivanov}}, \bibinfo {author} {\bibfnamefont {N.~E.}\ \bibnamefont {Kulagin}}, \ and\ \bibinfo {author} {\bibfnamefont {L.~M.}\ \bibnamefont {Lerman}},\ }\href {\doibase 10.1134/S002136401919007X} {\bibfield  {journal} {\bibinfo  {journal} {JETP Lett.}\ }\textbf {\bibinfo {volume} {110}},\ \bibinfo {pages} {481} (\bibinfo {year} {2019})}\BibitemShut {NoStop}%
\bibitem [{\citenamefont {Stanciu}\ \emph {et~al.}(2006)\citenamefont {Stanciu}, \citenamefont {Kimel}, \citenamefont {Hansteen}, \citenamefont {Tsukamoto}, \citenamefont {Itoh}, \citenamefont {Kirilyuk},\ and\ \citenamefont {Rasing}}]{stanciu2006ultrafast}%
  \BibitemOpen
  \bibfield  {author} {\bibinfo {author} {\bibfnamefont {C.~D.}\ \bibnamefont {Stanciu}}, \bibinfo {author} {\bibfnamefont {A.~V.}\ \bibnamefont {Kimel}}, \bibinfo {author} {\bibfnamefont {F.}~\bibnamefont {Hansteen}}, \bibinfo {author} {\bibfnamefont {A.}~\bibnamefont {Tsukamoto}}, \bibinfo {author} {\bibfnamefont {A.}~\bibnamefont {Itoh}}, \bibinfo {author} {\bibfnamefont {A.}~\bibnamefont {Kirilyuk}}, \ and\ \bibinfo {author} {\bibfnamefont {T.}~\bibnamefont {Rasing}},\ }\href {\doibase 10.1103/PhysRevB.73.220402} {\bibfield  {journal} {\bibinfo  {journal} {Phys. Rev. B}\ }\textbf {\bibinfo {volume} {73}},\ \bibinfo {pages} {220402} (\bibinfo {year} {2006})}\BibitemShut {NoStop}%
\bibitem [{\citenamefont {Ostler}\ \emph {et~al.}(2011)\citenamefont {Ostler}, \citenamefont {Evans}, \citenamefont {Chantrell}, \citenamefont {Atxitia}, \citenamefont {{Chubykalo-Fesenko}}, \citenamefont {Radu}, \citenamefont {Abrudan}, \citenamefont {Radu}, \citenamefont {Tsukamoto}, \citenamefont {Itoh}, \citenamefont {Kirilyuk}, \citenamefont {Rasing},\ and\ \citenamefont {Kimel}}]{ostler2011crystallographically}%
  \BibitemOpen
  \bibfield  {author} {\bibinfo {author} {\bibfnamefont {T.~A.}\ \bibnamefont {Ostler}}, \bibinfo {author} {\bibfnamefont {R.~F.~L.}\ \bibnamefont {Evans}}, \bibinfo {author} {\bibfnamefont {R.~W.}\ \bibnamefont {Chantrell}}, \bibinfo {author} {\bibfnamefont {U.}~\bibnamefont {Atxitia}}, \bibinfo {author} {\bibfnamefont {O.}~\bibnamefont {{Chubykalo-Fesenko}}}, \bibinfo {author} {\bibfnamefont {I.}~\bibnamefont {Radu}}, \bibinfo {author} {\bibfnamefont {R.}~\bibnamefont {Abrudan}}, \bibinfo {author} {\bibfnamefont {F.}~\bibnamefont {Radu}}, \bibinfo {author} {\bibfnamefont {A.}~\bibnamefont {Tsukamoto}}, \bibinfo {author} {\bibfnamefont {A.}~\bibnamefont {Itoh}}, \bibinfo {author} {\bibfnamefont {A.}~\bibnamefont {Kirilyuk}}, \bibinfo {author} {\bibfnamefont {T.}~\bibnamefont {Rasing}}, \ and\ \bibinfo {author} {\bibfnamefont {A.}~\bibnamefont {Kimel}},\ }\href {\doibase 10.1103/PhysRevB.84.024407} {\bibfield  {journal} {\bibinfo  {journal} {Phys. Rev. B}\ }\textbf {\bibinfo {volume} {84}},\ \bibinfo {pages}
  {024407} (\bibinfo {year} {2011})}\BibitemShut {NoStop}%
\bibitem [{\citenamefont {Kato}\ \emph {et~al.}(2008)\citenamefont {Kato}, \citenamefont {Nakazawa}, \citenamefont {Komiya}, \citenamefont {Nishizawa}, \citenamefont {Tsunashima},\ and\ \citenamefont {Iwata}}]{kato2008compositional}%
  \BibitemOpen
  \bibfield  {author} {\bibinfo {author} {\bibfnamefont {T.}~\bibnamefont {Kato}}, \bibinfo {author} {\bibfnamefont {K.}~\bibnamefont {Nakazawa}}, \bibinfo {author} {\bibfnamefont {R.}~\bibnamefont {Komiya}}, \bibinfo {author} {\bibfnamefont {N.}~\bibnamefont {Nishizawa}}, \bibinfo {author} {\bibfnamefont {S.}~\bibnamefont {Tsunashima}}, \ and\ \bibinfo {author} {\bibfnamefont {S.}~\bibnamefont {Iwata}},\ }\href {\doibase 10.1109/TMAG.2008.2001679} {\bibfield  {journal} {\bibinfo  {journal} {IEEE Transactions on Magnetics}\ }\textbf {\bibinfo {volume} {44}},\ \bibinfo {pages} {3380} (\bibinfo {year} {2008})}\BibitemShut {NoStop}%
\bibitem [{\citenamefont {Bainsla}\ \emph {et~al.}(2022)\citenamefont {Bainsla}, \citenamefont {Kumar}, \citenamefont {Awad}, \citenamefont {Wang}, \citenamefont {Zahedinejad}, \citenamefont {Behera}, \citenamefont {Fulara}, \citenamefont {Khymyn}, \citenamefont {Houshang}, \citenamefont {Weissenrieder},\ and\ \citenamefont {{\AA}kerman}}]{bainsla2022ultrathin}%
  \BibitemOpen
  \bibfield  {author} {\bibinfo {author} {\bibfnamefont {L.}~\bibnamefont {Bainsla}}, \bibinfo {author} {\bibfnamefont {A.}~\bibnamefont {Kumar}}, \bibinfo {author} {\bibfnamefont {A.~A.}\ \bibnamefont {Awad}}, \bibinfo {author} {\bibfnamefont {C.}~\bibnamefont {Wang}}, \bibinfo {author} {\bibfnamefont {M.}~\bibnamefont {Zahedinejad}}, \bibinfo {author} {\bibfnamefont {N.}~\bibnamefont {Behera}}, \bibinfo {author} {\bibfnamefont {H.}~\bibnamefont {Fulara}}, \bibinfo {author} {\bibfnamefont {R.}~\bibnamefont {Khymyn}}, \bibinfo {author} {\bibfnamefont {A.}~\bibnamefont {Houshang}}, \bibinfo {author} {\bibfnamefont {J.}~\bibnamefont {Weissenrieder}}, \ and\ \bibinfo {author} {\bibfnamefont {J.}~\bibnamefont {{\AA}kerman}},\ }\href {\doibase 10.1002/adfm.202111693} {\bibfield  {journal} {\bibinfo  {journal} {Advanced Functional Materials}\ }\textbf {\bibinfo {volume} {32}},\ \bibinfo {pages} {2111693} (\bibinfo {year} {2022})}\BibitemShut {NoStop}%
\bibitem [{\citenamefont {Kim}\ \emph {et~al.}(2017)\citenamefont {Kim}, \citenamefont {Kim}, \citenamefont {Hirata}, \citenamefont {Oh}, \citenamefont {Tono}, \citenamefont {Kim}, \citenamefont {Okuno}, \citenamefont {Ham}, \citenamefont {Kim}, \citenamefont {Go}, \citenamefont {Tserkovnyak}, \citenamefont {Tsukamoto}, \citenamefont {Moriyama}, \citenamefont {Lee},\ and\ \citenamefont {Ono}}]{kim2017fast}%
  \BibitemOpen
  \bibfield  {author} {\bibinfo {author} {\bibfnamefont {K.-J.}\ \bibnamefont {Kim}}, \bibinfo {author} {\bibfnamefont {S.~K.}\ \bibnamefont {Kim}}, \bibinfo {author} {\bibfnamefont {Y.}~\bibnamefont {Hirata}}, \bibinfo {author} {\bibfnamefont {S.-H.}\ \bibnamefont {Oh}}, \bibinfo {author} {\bibfnamefont {T.}~\bibnamefont {Tono}}, \bibinfo {author} {\bibfnamefont {D.-H.}\ \bibnamefont {Kim}}, \bibinfo {author} {\bibfnamefont {T.}~\bibnamefont {Okuno}}, \bibinfo {author} {\bibfnamefont {W.~S.}\ \bibnamefont {Ham}}, \bibinfo {author} {\bibfnamefont {S.}~\bibnamefont {Kim}}, \bibinfo {author} {\bibfnamefont {G.}~\bibnamefont {Go}}, \bibinfo {author} {\bibfnamefont {Y.}~\bibnamefont {Tserkovnyak}}, \bibinfo {author} {\bibfnamefont {A.}~\bibnamefont {Tsukamoto}}, \bibinfo {author} {\bibfnamefont {T.}~\bibnamefont {Moriyama}}, \bibinfo {author} {\bibfnamefont {K.-J.}\ \bibnamefont {Lee}}, \ and\ \bibinfo {author} {\bibfnamefont {T.}~\bibnamefont {Ono}},\ }\href {\doibase 10.1038/nmat4990} {\bibfield  {journal}
  {\bibinfo  {journal} {Nature Mater}\ }\textbf {\bibinfo {volume} {16}},\ \bibinfo {pages} {1187} (\bibinfo {year} {2017})}\BibitemShut {NoStop}%
\bibitem [{\citenamefont {Bl{\"a}sing}\ \emph {et~al.}(2018)\citenamefont {Bl{\"a}sing}, \citenamefont {Ma}, \citenamefont {Yang}, \citenamefont {Garg}, \citenamefont {Dejene}, \citenamefont {N'Diaye}, \citenamefont {Chen}, \citenamefont {Liu},\ and\ \citenamefont {Parkin}}]{blasing2018exchange}%
  \BibitemOpen
  \bibfield  {author} {\bibinfo {author} {\bibfnamefont {R.}~\bibnamefont {Bl{\"a}sing}}, \bibinfo {author} {\bibfnamefont {T.}~\bibnamefont {Ma}}, \bibinfo {author} {\bibfnamefont {S.-H.}\ \bibnamefont {Yang}}, \bibinfo {author} {\bibfnamefont {C.}~\bibnamefont {Garg}}, \bibinfo {author} {\bibfnamefont {F.~K.}\ \bibnamefont {Dejene}}, \bibinfo {author} {\bibfnamefont {A.~T.}\ \bibnamefont {N'Diaye}}, \bibinfo {author} {\bibfnamefont {G.}~\bibnamefont {Chen}}, \bibinfo {author} {\bibfnamefont {K.}~\bibnamefont {Liu}}, \ and\ \bibinfo {author} {\bibfnamefont {S.~S.~P.}\ \bibnamefont {Parkin}},\ }\href {\doibase 10.1038/s41467-018-07373-w} {\bibfield  {journal} {\bibinfo  {journal} {Nat Commun}\ }\textbf {\bibinfo {volume} {9}},\ \bibinfo {pages} {4984} (\bibinfo {year} {2018})}\BibitemShut {NoStop}%
\bibitem [{\citenamefont {Cai}\ \emph {et~al.}(2020)\citenamefont {Cai}, \citenamefont {Zhu}, \citenamefont {Lee}, \citenamefont {Mishra}, \citenamefont {Ren}, \citenamefont {Pollard}, \citenamefont {He}, \citenamefont {Liang}, \citenamefont {Teo},\ and\ \citenamefont {Yang}}]{cai2020ultrafast}%
  \BibitemOpen
  \bibfield  {author} {\bibinfo {author} {\bibfnamefont {K.}~\bibnamefont {Cai}}, \bibinfo {author} {\bibfnamefont {Z.}~\bibnamefont {Zhu}}, \bibinfo {author} {\bibfnamefont {J.~M.}\ \bibnamefont {Lee}}, \bibinfo {author} {\bibfnamefont {R.}~\bibnamefont {Mishra}}, \bibinfo {author} {\bibfnamefont {L.}~\bibnamefont {Ren}}, \bibinfo {author} {\bibfnamefont {S.~D.}\ \bibnamefont {Pollard}}, \bibinfo {author} {\bibfnamefont {P.}~\bibnamefont {He}}, \bibinfo {author} {\bibfnamefont {G.}~\bibnamefont {Liang}}, \bibinfo {author} {\bibfnamefont {K.~L.}\ \bibnamefont {Teo}}, \ and\ \bibinfo {author} {\bibfnamefont {H.}~\bibnamefont {Yang}},\ }\href {\doibase 10.1038/s41928-019-0345-8} {\bibfield  {journal} {\bibinfo  {journal} {Nat Electron}\ }\textbf {\bibinfo {volume} {3}},\ \bibinfo {pages} {37} (\bibinfo {year} {2020})}\BibitemShut {NoStop}%
\bibitem [{\citenamefont {Siddiqui}\ \emph {et~al.}(2018)\citenamefont {Siddiqui}, \citenamefont {Han}, \citenamefont {Finley}, \citenamefont {Ross},\ and\ \citenamefont {Liu}}]{siddiqui2018currentinduceda}%
  \BibitemOpen
  \bibfield  {author} {\bibinfo {author} {\bibfnamefont {S.~A.}\ \bibnamefont {Siddiqui}}, \bibinfo {author} {\bibfnamefont {J.}~\bibnamefont {Han}}, \bibinfo {author} {\bibfnamefont {J.~T.}\ \bibnamefont {Finley}}, \bibinfo {author} {\bibfnamefont {C.~A.}\ \bibnamefont {Ross}}, \ and\ \bibinfo {author} {\bibfnamefont {L.}~\bibnamefont {Liu}},\ }\href {\doibase 10.1103/PhysRevLett.121.057701} {\bibfield  {journal} {\bibinfo  {journal} {Phys. Rev. Lett.}\ }\textbf {\bibinfo {volume} {121}},\ \bibinfo {pages} {057701} (\bibinfo {year} {2018})}\BibitemShut {NoStop}%
\bibitem [{\citenamefont {Caretta}\ \emph {et~al.}(2018)\citenamefont {Caretta}, \citenamefont {Mann}, \citenamefont {B{\"u}ttner}, \citenamefont {Ueda}, \citenamefont {Pfau}, \citenamefont {G{\"u}nther}, \citenamefont {Hessing}, \citenamefont {Churikova}, \citenamefont {Klose}, \citenamefont {Schneider}, \citenamefont {Engel}, \citenamefont {Marcus}, \citenamefont {Bono}, \citenamefont {Bagschik}, \citenamefont {Eisebitt},\ and\ \citenamefont {Beach}}]{caretta2018fast}%
  \BibitemOpen
  \bibfield  {author} {\bibinfo {author} {\bibfnamefont {L.}~\bibnamefont {Caretta}}, \bibinfo {author} {\bibfnamefont {M.}~\bibnamefont {Mann}}, \bibinfo {author} {\bibfnamefont {F.}~\bibnamefont {B{\"u}ttner}}, \bibinfo {author} {\bibfnamefont {K.}~\bibnamefont {Ueda}}, \bibinfo {author} {\bibfnamefont {B.}~\bibnamefont {Pfau}}, \bibinfo {author} {\bibfnamefont {C.~M.}\ \bibnamefont {G{\"u}nther}}, \bibinfo {author} {\bibfnamefont {P.}~\bibnamefont {Hessing}}, \bibinfo {author} {\bibfnamefont {A.}~\bibnamefont {Churikova}}, \bibinfo {author} {\bibfnamefont {C.}~\bibnamefont {Klose}}, \bibinfo {author} {\bibfnamefont {M.}~\bibnamefont {Schneider}}, \bibinfo {author} {\bibfnamefont {D.}~\bibnamefont {Engel}}, \bibinfo {author} {\bibfnamefont {C.}~\bibnamefont {Marcus}}, \bibinfo {author} {\bibfnamefont {D.}~\bibnamefont {Bono}}, \bibinfo {author} {\bibfnamefont {K.}~\bibnamefont {Bagschik}}, \bibinfo {author} {\bibfnamefont {S.}~\bibnamefont {Eisebitt}}, \ and\ \bibinfo {author} {\bibfnamefont {G.~S.~D.}\
  \bibnamefont {Beach}},\ }\href {\doibase 10.1038/s41565-018-0255-3} {\bibfield  {journal} {\bibinfo  {journal} {Nature Nanotech}\ }\textbf {\bibinfo {volume} {13}},\ \bibinfo {pages} {1154} (\bibinfo {year} {2018})}\BibitemShut {NoStop}%
\bibitem [{\citenamefont {Ghosh}\ \emph {et~al.}(2021)\citenamefont {Ghosh}, \citenamefont {Komori}, \citenamefont {Hallal}, \citenamefont {Pe{\~n}a~Garcia}, \citenamefont {Gushi}, \citenamefont {Hirose}, \citenamefont {Mitarai}, \citenamefont {Okuno}, \citenamefont {Vogel}, \citenamefont {Chshiev}, \citenamefont {Attan{\'e}}, \citenamefont {Vila}, \citenamefont {Suemasu},\ and\ \citenamefont {Pizzini}}]{ghosh2021currentdriven}%
  \BibitemOpen
  \bibfield  {author} {\bibinfo {author} {\bibfnamefont {S.}~\bibnamefont {Ghosh}}, \bibinfo {author} {\bibfnamefont {T.}~\bibnamefont {Komori}}, \bibinfo {author} {\bibfnamefont {A.}~\bibnamefont {Hallal}}, \bibinfo {author} {\bibfnamefont {J.}~\bibnamefont {Pe{\~n}a~Garcia}}, \bibinfo {author} {\bibfnamefont {T.}~\bibnamefont {Gushi}}, \bibinfo {author} {\bibfnamefont {T.}~\bibnamefont {Hirose}}, \bibinfo {author} {\bibfnamefont {H.}~\bibnamefont {Mitarai}}, \bibinfo {author} {\bibfnamefont {H.}~\bibnamefont {Okuno}}, \bibinfo {author} {\bibfnamefont {J.}~\bibnamefont {Vogel}}, \bibinfo {author} {\bibfnamefont {M.}~\bibnamefont {Chshiev}}, \bibinfo {author} {\bibfnamefont {J.-P.}\ \bibnamefont {Attan{\'e}}}, \bibinfo {author} {\bibfnamefont {L.}~\bibnamefont {Vila}}, \bibinfo {author} {\bibfnamefont {T.}~\bibnamefont {Suemasu}}, \ and\ \bibinfo {author} {\bibfnamefont {S.}~\bibnamefont {Pizzini}},\ }\href {\doibase 10.1021/acs.nanolett.1c00125} {\bibfield  {journal} {\bibinfo  {journal} {Nano Lett.}\
  }\textbf {\bibinfo {volume} {21}},\ \bibinfo {pages} {2580} (\bibinfo {year} {2021})}\BibitemShut {NoStop}%
\bibitem [{\citenamefont {Okuno}(2020)}]{okuno2020spintransfer}%
  \BibitemOpen
  \bibfield  {author} {\bibinfo {author} {\bibfnamefont {T.}~\bibnamefont {Okuno}},\ }in\ \href {\doibase 10.1007/978-981-15-9176-1_2} {\emph {\bibinfo {booktitle} {Magnetic {{Dynamics}} in {{Antiferromagnetically-Coupled Ferrimagnets}}: {{The Role}} of {{Angular Momentum}}}}},\ \bibinfo {series and number} {Springer {{Theses}}},\ \bibinfo {editor} {edited by\ \bibinfo {editor} {\bibfnamefont {T.}~\bibnamefont {Okuno}}}\ (\bibinfo  {publisher} {{Springer}},\ \bibinfo {address} {{Singapore}},\ \bibinfo {year} {2020})\ pp.\ \bibinfo {pages} {25--48}\BibitemShut {NoStop}%
\bibitem [{\citenamefont {Zvezdin}\ \emph {et~al.}(2020)\citenamefont {Zvezdin}, \citenamefont {Gareeva},\ and\ \citenamefont {Zvezdin}}]{zvezdin2020anomalies}%
  \BibitemOpen
  \bibfield  {author} {\bibinfo {author} {\bibfnamefont {A.~K.}\ \bibnamefont {Zvezdin}}, \bibinfo {author} {\bibfnamefont {Z.~V.}\ \bibnamefont {Gareeva}}, \ and\ \bibinfo {author} {\bibfnamefont {K.~A.}\ \bibnamefont {Zvezdin}},\ }\href {\doibase 10.1016/j.jmmm.2020.166876} {\bibfield  {journal} {\bibinfo  {journal} {Journal of Magnetism and Magnetic Materials}\ }\textbf {\bibinfo {volume} {509}},\ \bibinfo {pages} {166876} (\bibinfo {year} {2020})}\BibitemShut {NoStop}%
\bibitem [{\citenamefont {Logunov}\ \emph {et~al.}(2021)\citenamefont {Logunov}, \citenamefont {Safonov}, \citenamefont {Fedorov}, \citenamefont {Danilova}, \citenamefont {Moiseev}, \citenamefont {Safin}, \citenamefont {Nikitov},\ and\ \citenamefont {Kirilyuk}}]{logunov2021domain}%
  \BibitemOpen
  \bibfield  {author} {\bibinfo {author} {\bibfnamefont {M.}~\bibnamefont {Logunov}}, \bibinfo {author} {\bibfnamefont {S.}~\bibnamefont {Safonov}}, \bibinfo {author} {\bibfnamefont {A.}~\bibnamefont {Fedorov}}, \bibinfo {author} {\bibfnamefont {A.}~\bibnamefont {Danilova}}, \bibinfo {author} {\bibfnamefont {N.}~\bibnamefont {Moiseev}}, \bibinfo {author} {\bibfnamefont {A.}~\bibnamefont {Safin}}, \bibinfo {author} {\bibfnamefont {S.}~\bibnamefont {Nikitov}}, \ and\ \bibinfo {author} {\bibfnamefont {A.}~\bibnamefont {Kirilyuk}},\ }\href {\doibase 10.1103/PhysRevApplied.15.064024} {\bibfield  {journal} {\bibinfo  {journal} {Phys. Rev. Appl.}\ }\textbf {\bibinfo {volume} {15}},\ \bibinfo {pages} {064024} (\bibinfo {year} {2021})}\BibitemShut {NoStop}%
\bibitem [{\citenamefont {Galkina}\ \emph {et~al.}(2022)\citenamefont {Galkina}, \citenamefont {Kireev},\ and\ \citenamefont {Ivanov}}]{galkina2022solitons}%
  \BibitemOpen
  \bibfield  {author} {\bibinfo {author} {\bibfnamefont {E.~G.}\ \bibnamefont {Galkina}}, \bibinfo {author} {\bibfnamefont {V.~E.}\ \bibnamefont {Kireev}}, \ and\ \bibinfo {author} {\bibfnamefont {B.~A.}\ \bibnamefont {Ivanov}},\ }\href {\doibase 10.1063/10.0014580} {\bibfield  {journal} {\bibinfo  {journal} {Low Temperature Physics}\ }\textbf {\bibinfo {volume} {48}},\ \bibinfo {pages} {896} (\bibinfo {year} {2022})}\BibitemShut {NoStop}%
\bibitem [{\citenamefont {Dillon}(1963)}]{dillon1963domains}%
  \BibitemOpen
  \bibfield  {author} {\bibinfo {author} {\bibfnamefont {J.~F.}\ \bibnamefont {Dillon}},\ }in\ \href {\doibase 10.1016/B978-0-12-575303-6.50016-6} {\emph {\bibinfo {booktitle} {Spin {{Arrangements}} and {{Crystal Structure}}, {{Domains}}, and {{Micromagnetics}}}}},\ \bibinfo {editor} {edited by\ \bibinfo {editor} {\bibfnamefont {G.~T.}\ \bibnamefont {Rado}}\ and\ \bibinfo {editor} {\bibfnamefont {H.}~\bibnamefont {Suhl}}}\ (\bibinfo  {publisher} {{Academic Press}},\ \bibinfo {year} {1963})\ pp.\ \bibinfo {pages} {415--464}\BibitemShut {NoStop}%
\bibitem [{\citenamefont {Ivanov}\ and\ \citenamefont {Sukstanskii}(1983)}]{ivanov1983nonlinear}%
  \BibitemOpen
  \bibfield  {author} {\bibinfo {author} {\bibfnamefont {B.~A.}\ \bibnamefont {Ivanov}}\ and\ \bibinfo {author} {\bibfnamefont {A.~L.}\ \bibnamefont {Sukstanskii}},\ }\href@noop {} {\bibfield  {journal} {\bibinfo  {journal} {Zhurnal Eksperimentalnoi i Teoreticheskoi Fiziki}\ }\textbf {\bibinfo {volume} {84}},\ \bibinfo {pages} {370} (\bibinfo {year} {1983})}\BibitemShut {NoStop}%
\bibitem [{\citenamefont {Ivanov}\ \emph {et~al.}(2020)\citenamefont {Ivanov}, \citenamefont {Galkina}, \citenamefont {Kireev}, \citenamefont {Kulagin}, \citenamefont {Ovcharov},\ and\ \citenamefont {Khymyn}}]{ivanov2020nonstationary}%
  \BibitemOpen
  \bibfield  {author} {\bibinfo {author} {\bibfnamefont {B.~A.}\ \bibnamefont {Ivanov}}, \bibinfo {author} {\bibfnamefont {E.~G.}\ \bibnamefont {Galkina}}, \bibinfo {author} {\bibfnamefont {V.~E.}\ \bibnamefont {Kireev}}, \bibinfo {author} {\bibfnamefont {N.~E.}\ \bibnamefont {Kulagin}}, \bibinfo {author} {\bibfnamefont {R.~V.}\ \bibnamefont {Ovcharov}}, \ and\ \bibinfo {author} {\bibfnamefont {R.~S.}\ \bibnamefont {Khymyn}},\ }\href {\doibase 10.1063/10.0001552} {\bibfield  {journal} {\bibinfo  {journal} {Low Temperature Physics}\ }\textbf {\bibinfo {volume} {46}},\ \bibinfo {pages} {841} (\bibinfo {year} {2020})}\BibitemShut {NoStop}%
\bibitem [{\citenamefont {Vansteenkiste}\ \emph {et~al.}(2014)\citenamefont {Vansteenkiste}, \citenamefont {Leliaert}, \citenamefont {Dvornik}, \citenamefont {Helsen}, \citenamefont {{Garcia-Sanchez}},\ and\ \citenamefont {Van~Waeyenberge}}]{vansteenkiste2014designa}%
  \BibitemOpen
  \bibfield  {author} {\bibinfo {author} {\bibfnamefont {A.}~\bibnamefont {Vansteenkiste}}, \bibinfo {author} {\bibfnamefont {J.}~\bibnamefont {Leliaert}}, \bibinfo {author} {\bibfnamefont {M.}~\bibnamefont {Dvornik}}, \bibinfo {author} {\bibfnamefont {M.}~\bibnamefont {Helsen}}, \bibinfo {author} {\bibfnamefont {F.}~\bibnamefont {{Garcia-Sanchez}}}, \ and\ \bibinfo {author} {\bibfnamefont {B.}~\bibnamefont {Van~Waeyenberge}},\ }\href {\doibase 10.1063/1.4899186} {\bibfield  {journal} {\bibinfo  {journal} {AIP Advances}\ }\textbf {\bibinfo {volume} {4}},\ \bibinfo {pages} {107133} (\bibinfo {year} {2014})}\BibitemShut {NoStop}%
\bibitem [{sup()}]{supplemental}%
  \BibitemOpen
  \href@noop {} {}\bibinfo {howpublished} {[URL will be inserted by a publisher]}\BibitemShut {NoStop}%
\end{thebibliography}%

\end{document}

\title[]{Supplementary material for \\Instability in domain wall dynamics in almost compensated ferrimagnets}

\author{R. V. Ovcharov}
\affiliation{ 
Department of Physics, University of Gothenburg, Gothenburg 41296, Sweden
}

\author{B. A. Ivanov}
\affiliation{
Institute of Magnetism of NASU and MESU, Kyiv 03142, Ukraine
}
\affiliation{
William H. Miller III Department of Physics and Astronomy, Johns Hopkins University, Baltimore, MD 21218, USA.
}

\author{E. G. Galkina}%
\affiliation{ 
Institute of Physics, National Academy of Sciences of Ukraine, Kyiv, 03028, Ukraine
}

\author{J. \AA kerman}
\affiliation{ 
Department of Physics, University of Gothenburg, Gothenburg 41296, Sweden
}
\affiliation{
Center for Science and Innovation in Spintronics, Tohoku University, Sendai 980-8577, Japan
}
\affiliation{
Research Institute of Electrical Communication, Tohoku University, Sendai 980-8577, Japan
}

\author{R. S.  Khymyn}%
\affiliation{ 
Department of Physics, University of Gothenburg, Gothenburg 41296, Sweden
}

\maketitle

The ferrimagnetic structure is modeled by separate regions defined for each magnetic sublattice, similar to what was proposed in Ref.~[5]. However, in our framework, the magnetization in the simulation represents a scaled spin density so that it coincides with one of the magnetizations of the ferrimagnet $\mathbf{M}=-\gamma \hbar \mathbf{S}$, where gyromagnetic ratio $\gamma$ is the \emph{MuMax3} global parameter. At the spin compensation point, $\mathbf{M}$ in both layers are equal by magnitude. Spin imbalance is realized by changing parameter $\mathbf{M}$ in different layers so that their sum remains unchanged. This representation avoids the renormalization of material parameters for distinct layers. The difference in $g$-factors was only considered for the external field; a magnetic field applied to one of the magnetic sublattices is scaled by a factor of $g_{FeCo}(=2.2)/g_{Gd}(=2.0)=1.1$. The strip is initialized with the DW located near the edge. The external field is applied along the easy axis of the FiM. The DW position is determined by the element with the zero $z$-projection of the N\'eel vector, and the angle $\phi$ is calculated at this point.

For stationary motion $v=const$, the dependence of the established velocity on the applied force $v(F)$ is found from the balance between friction $F_{fr}(v)$  and driving $F$ forces. For small values of velocity ($v \ll c$) $F_{fr}\propto v$, there is a linear dependence on the driving force:
\begin{equation}
    \label{eq:velocityOnForceLinear}
     v =\mu F,  \quad \mu = l_0 / 2 \alpha \hbar S_{tot},
\end{equation}
where $\mu$ means mobility for the specific type of the DW. With an increase in the driving force, the velocity dependence becomes more complicated and differs significantly depending on the limiting speed, which, in turn, depends on the imbalance parameter.\\

Precisely at the compensation point, the limiting speed is $v_{c} = c$; as the velocity reaches this limit, the friction force $F_{fr}(v)$ increases indefinitely. For the inverse dependence $v(F)$, the DW velocity monotonically tends to $c$ as the driving force increases:
\begin{equation}
    \label{eq:velocityOnForceAFM} 
    v(F) = \frac{\mu F c}{\sqrt{(\mu F)^2 + c^2}} 
\end{equation}

The results of micromagnetic simulations confirming this dependence are shown in Fig.~\ref{img:velocituVsForce}. The wall motion is shown in Supplementary Video S1, the FiM DW at the spin compensation point behaves similarly to the AFM one. The only dynamical change of the DW occurs during the transition to a stationary regime and is associated with the DW contraction. The internal angle $\phi$ remains at its initial value regardless of the DW velocity.\\

\begin{figure}[hbt!]
\includegraphics[width=0.55\textwidth]{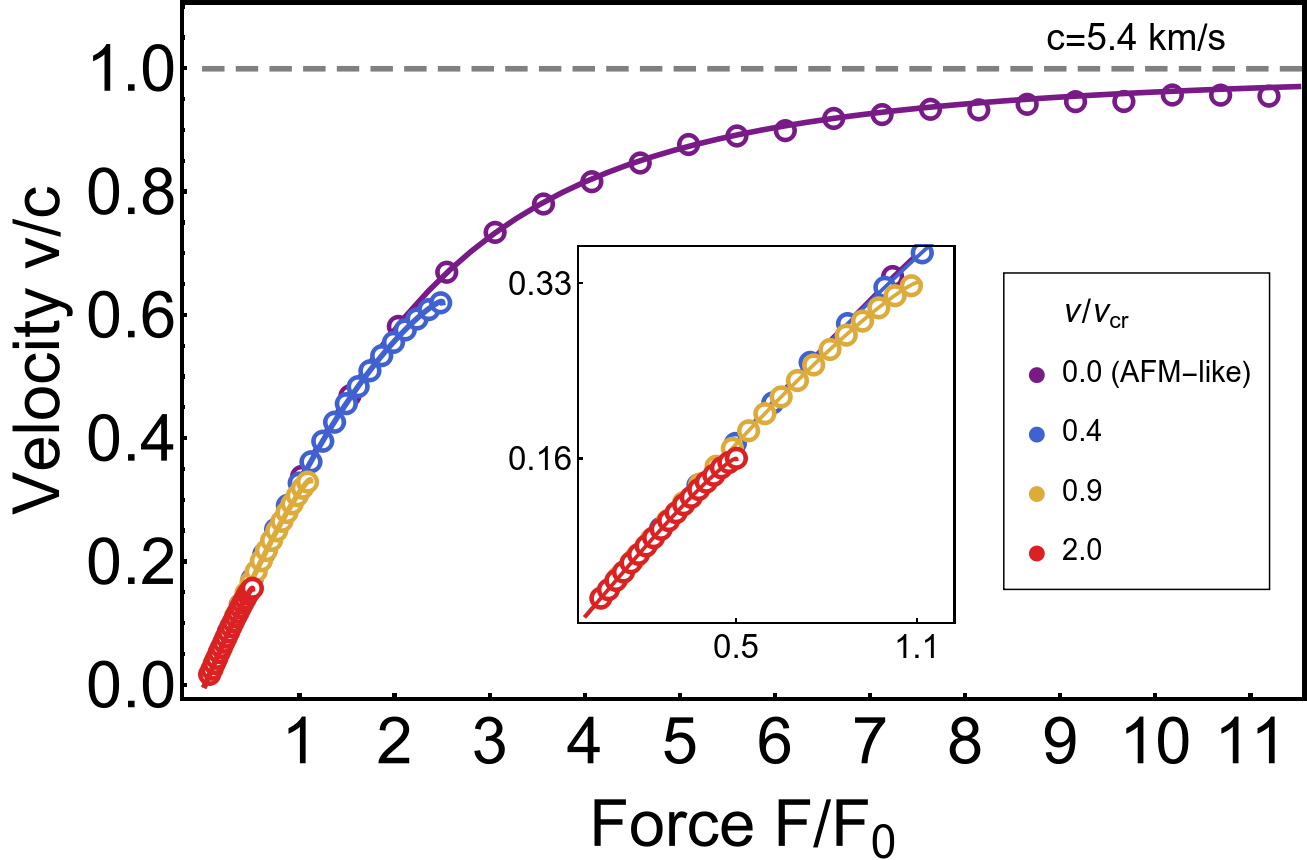}
 \caption{\textbf{Stationary motion.} The DW velocity as a function of applied force for different spin imbalances. The normalization force $F_{0}$ corresponds to the maximum friction force $F_{max}$ for $\nu=\nu_{cr}$. Circles show the results of micromagnetic simulations, while solid lines are calculated analytically using Eqs.~(\ref{eq:velocityOnForceAFM}, \ref{eq:velocityOnForceFiM}). }
\label{img:velocituVsForce}
\end{figure}

For non-zero imbalance $\nu \neq 0$, the friction force is limited from above $F_{fr}(v) \leq F_{max}$. Stationary motion is realized at $F\leq F_{max}$ with the set velocity:
\begin{equation}
    \label{eq:velocityOnForceFiM}
    v(F) = \frac{\mu F c}{\sqrt{(\mu F)^2 + (c^2/2)(2 + \rho \mp \rho\sqrt{1-F^2/F_{max}^2})}} 
\end{equation}

Here, the signs ``$-$'' and ``$+$'' correspond to the Bloch and N\'eel walls, respectively. The motion with constant velocity corresponds to the particular point on the dispersion dependence. As driving force balances the maximum friction force $F_{max}$, the DW will reach its highest momentum of steady-state motion $P_{max}$. In the important case of $\nu < \nu_{cr}$, the point with $P_{max}$ is placed before the characteristic region of the dispersion endpoint, allowing stable stationary motion even at $\nu < \nu_{cr}$. Supplementary Video S2 shows it for the case of $\nu = 0.4\nu_{cr}$ with the applied force slightly below $F_{max}$. The DW width experiences a minor change since the velocity is far from the relativistic limit $c$. \newpage Here, the prominent feature of DW dynamics is the transition of the internal angle $\phi$ to the stationary value. The results of micromagnetic simulations verifying dependence (\ref{eq:velocityOnForceFiM}) for different values of spin imbalance are shown in Fig.~\ref{img:velocituVsForce}.\\

The non-stationary motion is achieved at $F > F_{max}$. When the imbalance is beyond the critical value, the DW behaves like an FM one, remaining stable during movement. This behavior is verified through simulations with $\nu=2\nu_{cr}$. Supplementary video S3 demonstrates the mechanism of non-stationary motion with the applied force of $2F_{max}$. The orientation of the spins within the wall, i.e., angle $\phi$, constantly changes; that is, Walker's supercritical motion is realized. With the slow oscillations of the angle $\phi$ , the velocity changes, leading to a temporary movement of the wall in the opposite direction following Eq.~(3b) given in the main text.

\begin{figure}[hbt!]
\includegraphics[width=0.55\textwidth]{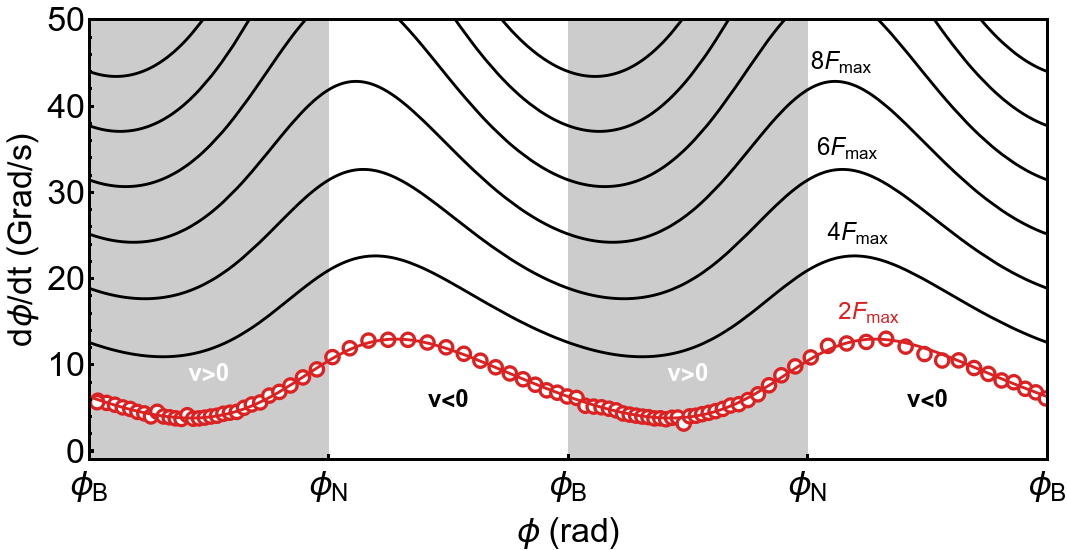}
 \caption{\textbf{FM-like non-stationary motion.} DW instantaneous frequency $d_t \phi$ \textit{vs.} angle $\phi$ over time for different values of the driving force. Spin imbalance $\nu=2\nu_{cr}$. Circles show the results of micromagnetic simulations, while lines are built according to the numerical solution of Eq.~(\ref{eq:velocityOnForceFiM}). The areas colored in gray indicate that the wall moves with a positive velocity $v>0$ according to Eq.~(3b), while for white areas $v<0$. }
\label{img:phaseAndFrequency}
\end{figure}

To analyze the non-stationary motion of a DW at $\nu > \nu_{cr}$, the reduced form of the equation for $dP/dt$ (Eq.~1 in the main text) can be used~[42]:
\begin{equation}
\label{eq:dpdtInPhi}
    \frac{d \phi}{dt}\left[ 1 + \left( \frac{\nu_{cr}}{\nu} \right)^2 \cos 2\phi \right] = \alpha \omega_{ex}\frac{1}{2} \left( \frac{\nu_{cr}}{\nu} \right)^2 \left[ \frac{F}{F_{max}} - \sin 2 \phi \right]
\end{equation}

The solution to this equation allows one to determine the velocity using Eq.~(3b), and thus the coordinate of the wall $X(t) = \int v(t) dt$. For spin imbalances $\nu>\nu_{cr}$, Eq.~(\ref{eq:dpdtInPhi}) can be solved numerically. The results of numerical calculations are in good agreement with the results of micromagnetic simulations, see Fig.~\ref{img:phaseAndFrequency}. At forces that slightly exceed $F_{max}$, the wall changes the angle $\phi$ more slowly at positive velocities (from $\phi_B$ to $\phi_N$) and passes through the region of negative velocities faster (from $\phi_N$ to $\phi_B$). Due to this, on average, the wall continues movement to one of the sides. With the further increase of the driving force, the phase change in these regions becomes similar, so the average velocity tends to zero. Equation~(\ref{eq:dpdtInPhi}), however, cannot be solved numerically for the case $\nu<\nu_{cr}$ and $F>F_{max}$, which was analyzed using micromagnetic simulations and is discussed in the main text.